\documentclass[letter]{aa}

\usepackage{graphicx}
\usepackage{epstopdf}
\usepackage{pdflscape}
\usepackage{booktabs}
\usepackage{txfonts}
\usepackage{gensymb}
\usepackage{flushend}

\let\xxtabular\tabular
\renewcommand{\tabular}{\small\xxtabular}

\def\Sn{S_{\nu}}
\def\RS{R_{\rm S}}
\def\LR{L_{\rm R}}
\def\Pj{P_{\rm jet}}
\def\Pacc{P_{\rm acc}}

\begin{document} 

\title{No asymmetric outflows from Sagittarius A* during\\ the pericenter passage of the gas cloud G2}

\author{J.-H. Park\inst{\ref{inst1}} \and S. Trippe\thanks{Corresponding author}\inst{\ref{inst1}} \and T. P. Krichbaum\inst{\ref{inst2}} \and J.-Y. Kim\inst{\ref{inst1}} \and M. Kino\inst{\ref{inst3}} \and A. Bertarini\inst{\ref{inst2}, \ref{inst4}} \and M. Bremer\inst{\ref{inst5}} \and P. de Vicente\inst{\ref{inst6}} }

\institute{Department of Physics and Astronomy, Seoul National University, 1 Gwanak-ro, Gwanak-gu, Seoul 151-742, Korea\\ \email{jhpark@astro.snu.ac.kr, trippe@astro.snu.ac.kr}\label{inst1}
\and Max-Planck-Institut f{\"u}r Radioastronomie (MPIfR), Auf dem H{\"u}gel 69, 53121 Bonn, Germany\label{inst2}
\and Korea Astronomy and Space Science Institute (KASI), 776 Daedeokdae-ro, Yuseong-gu, Daejeon 305-348, Korea\label{inst3}
\and Institute of Geodesy and Geoinformation, Bonn University, Nussallee 17, 53115 Bonn, Germany\label{inst4}
\and Institut de Radioastronomie Millim{\'e}trique (IRAM), 300 rue de la Piscine, 38406 Saint-Martin d'H\`eres, France\label{inst5}
\and Observatorio Astronomico Nacional (IGN), Observatorio de Yebes, Aptdo. 148, Yebes, 19080 Guadalajara, Spain\label{inst6} }

\date{Received ---; Accepted ---}

\abstract{
The gas cloud G2 falling toward Sagittarius A* (Sgr~A*), the supermassive black hole at the center of the Milky Way, is supposed to provide valuable information on the physics of accretion flows and the environment of the black hole. We observed Sgr~A* with four European stations of the Global Millimeter Very Long Baseline Interferometry Array (GMVA) at 86 GHz on 1 October 2013 when parts of G2 had already passed the pericenter. We searched for possible transient asymmetric structure -- such as jets or winds from hot accretion flows -- around Sgr~A* caused by accretion of material from G2. The interferometric closure phases remained zero within errors during the observation time. We thus conclude that Sgr~A* did not show significant asymmetric (in the observer frame) outflows in late 2013. Using simulations, we constrain the size of the outflows that we could have missed to $\approx$2.5 mas along the major axis, $\approx$0.4 mas along the minor axis of the beam, corresponding to approximately 232 and 35 Schwarzschild radii, respectively; we thus probe spatial scales on which the jets of radio galaxies are suspected to convert magnetic into kinetic energy. As probably less than 0.2~Jy of the flux from Sgr~A* can be attributed to accretion from G2, one finds an effective accretion rate $\eta\dot{M}\lesssim1.5\times10^9~{\rm kg\,s}^{-1}\approx7.7\times10^{-9}\,M_{\oplus}\,{\rm yr}^{-1}$ for material from G2. Exploiting the kinetic jet power--accretion power relation of radio galaxies, one finds that the rate of accretion of matter that ends up in jets is limited to $\dot{M}\lesssim10^{17}\,{\rm kg\,s}^{-1}\approx0.5M_{\oplus}\,{\rm yr}^{-1}$, less than about 20\% of the mass of G2. Accordingly, G2 appears to be largely stable against loss of angular momentum and subsequent (partial) accretion at least on time scales $\lesssim$1 year.
}
	

\titlerunning{No asymmetric outflows from Sgr~A* during the pericenter passage of G2}

\authorrunning{J.-H. Park et al.}

\maketitle


\section{Introduction}

With a mass of $M_{\bullet}\approx4.3\times10^6 M_{\odot}$ and located at the center of the Milky Way at a distance of $R_0\approx8$~kpc, Sagittarius~A* (Sgr~A*) is the nearest supermassive black hole (see, e.g., \citealt{Genzel2010} for a review). Thanks to its proximity, the Galactic center serves as an excellent laboratory for the astrophysics of galactic nuclei; for the given mass and distance, $1\,{\rm milliarcsecond}\equiv8\,{\rm a.u.}\equiv94\,\RS$, with $\RS$ denoting the Schwarzschild radius. Sgr~A* is characterized by its low luminosity compared to active galactic nuclei (AGN), with a bolometric luminosity $\lesssim2\times10^{-8}$ of its Eddington luminosity (see \citealt{Narayan1998} and references therein). Sgr~A* shows a slightly inverted radio spectrum peaking at mm-to-submm wavelengths \citep{Zylka1992, Falcke1998, Melia2001, Bower2015}.

The emission mechanism of Sgr~A* is a matter of ongoing debate. On the one hand, radiatively inefficient accretion flows (RIAFs; e.g., \citealt{Yuan2003}) reproduce the observed fluxes at various wavelengths successfully. On the other hand, multiple studies \citep{Falcke2000, Markoff2001, Yuan2002, Markoff2007, Falcke2009} have pointed out the need for parts of the emission to originate from jets, as well as observational evidence for the presence of jets \citep{Yusef2012,Li2013}. The current lack of structural information for Sgr~A* prevents an unambiguous decision between competing theories. At the short-wavelength side of the submm-bump, where the emission becomes optically thin, instruments with the necessary spatial resolution are not available. At wavelengths of more than a few millimeters, the source structure is washed out by interstellar scattering (e.g., \citealt{Bower2006}). \citet{Lu2011} found evidence that the shape and orientation of the elliptical Gaussian changes with frequency; this could be interpreted as an intrinsic structure which is slightly misaligned with the scattering disk, shining through towards higher frequencies. In addition, recent Very Long Baseline Interferometry (VLBI) observations at 43~GHz have been able to resolve an intrinsic elliptical structure with a preferred geometrical axis \citep{Bower2014}. This structure might indicate an existence of jets but could also be the result of an elongated accretion flow such as a black hole crescent (e.g., \citealt{Kamruddin2013}).

Recently, a gas cloud labeled G2 was observed to move toward Sgr~A* on a nearly radial orbit \citep{Gillessen2012}. As yet, two possible structures of G2 have been discussed mainly, with the first scenario being that G2 is a localized over-dense region within an extended gas streamer. This is in agreement with observations reporting that G2 is composed of a compact head and a more widespread tail \citep{Gillessen2013b, Pfuhl2015}. The two components are on approximately the same orbit and connected by a faint bridge in position-velocity diagrams, indicating that they might share the same origin. According to this scenario, the pericenter passage started in early 2013 \citep{Gillessen2013a} and lasted over one year while G2 has been stretched substantially along its orbit by tidal shearing caused by the gravitational potential of Sgr A* \citep{Pfuhl2015}. Test particle simulations have provided a good explanation on the dynamics of G2, corresponding that hydrodynamic effects have not been significant (e.g., \citealt{Pfuhl2015}; see also \citealt{Schartmann2012}). The second possibility under discussion is that G2 is a circumstellar cloud around a star which provides stabilizing gravity and replenishes gas continuously. Studies supporting this scenario pay attention to the possibility that the tail is just a fore/background feature and not physically connected to G2 \citep{Phifer2013, Valencia-S.2015}. Recent observations also have shown that G2 survived its pericenter passage as a compact source, which might be a clue for G2 being a star enshrouded by gas and/or dust \citep{Witzel2014, Valencia-S.2015}. This scenario was modeled analytically \citep{Miralda2012, Scoville2013} and explored with the help of numerical simulations \citep{Ballone2013, Colle2014, Zajacek2014}.

Interactions with the accretion flows toward Sgr~A* might cause G2 to lose angular momentum, resulting in parts of it to be accreted by the black hole \citep{Anninos2012, Burkert2012, Schartmann2012}. Accordingly, one may expect an increased radio luminosity (e.g., \citealt{Mahadevan1997, Moscibrodzka2012}) as well as an increase in source size (e.g., \citealt{Moscibrodzka2012}). In addition, radio-bright outflows, like jets or wind-like outflows related to RIAFs \citep{Yuan2003, Moscibrodzka2012, Liu2013}, might become observable on spatial scales of $\lesssim 1$ mas. In order to search for such transient structure, we performed VLBI observations with four European stations of the Global Millimeter VLBI Array (GMVA) at 86 GHz, providing an angular resolution down to $\approx$0.3~mas. Our observing frequency of 86~GHz is in a region where scattering vanishes and dominates less the images of Sgr~A* at frequencies less than about 43~GHz  \citep{Bower2004,Bower2006, Shen2005}.

\begin{figure}[!t]
\centering
\includegraphics[trim=4mm 13mm 7mm 5mm, clip, width = 89mm]{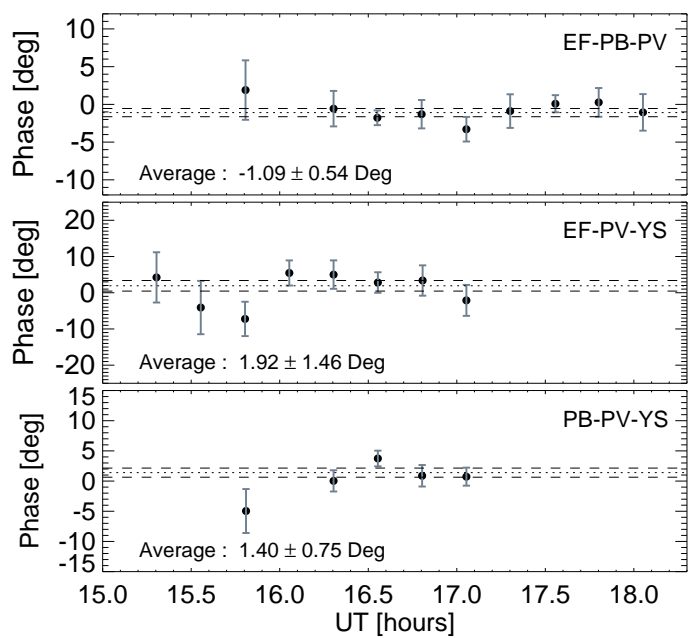}
\caption{Closure phases vs. time for three independent VLBI triangles. Each data point denotes a closure phase measurement averaged over one scan of 6 minutes; error bars correspond to $1\sigma$ errors. Dotted lines indicate weighted averages of all data points, dashed lines represent the corresponding standard errors of mean. Average phase values are noted in each diagram. Phases for the triangle EF-PB-PV are extracted from Stokes I data, phases for the other triangles are from LL data.}
\label{Closure}
\end{figure}

\section{Observations and Data Analysis}

We observed Sgr~A* at 86 GHz on 1 October 2013 using four GMVA stations of the European VLBI Network (EVN): Effelsberg (EF), Pico Veleta (PV), Plateau de Bure (PB), and Yebes (YS). A combination of low declination of Sgr~A* ($-29\degree$), high latitude of the stations (the minimum latitude being $37\degree$ for Pico Veleta), and a technical problem at Yebes station limited our observing times to around 1.5 hours for Yebes and 2.5 hours for the other stations.  We observed both circular polarizations except of Yebes station where only left circular polarization (LCP) data were obtained. The data were recorded with the Mark~5 VLBI system (2-bit sampling) using the digital baseband converter (DBBC) in polyphase filter bank mode with a bandwidth of 32 channels, each 16 MHz wide (total bandwidth 512 MHz). All data were correlated with the DifX VLBI correlator of the MPIfR (Bonn, Germany). Our calibrators (fringe finding sources) were NRAO~530 and 1633+38.

We followed the standard procedures for initial phase and amplitude calibration using the AIPS software package \citep{Greisen2003} and applied phase self-calibration using the Caltech Difmap package \citep{Shepherd1997} with a point-source model. The geometry of our observation resulted in a very elongated beam (point spread function) with full widths at half maximum (FWHM) of 3.02~mas and 0.33~mas, respectively, at a position angle of $-22\degree$. For the flux density of Sgr~A* we found a value of $\approx$1.4 $\pm$ 0.3~Jy).

We used the evolution of \emph{closure phases} to search for asymmetric, extended emission around Sgr~A*. A closure phase is the sum of the interferometric phases of the three baselines in a closed triangle of stations; it is free from antenna-based phase errors (e.g., \citealt{Thompson2001}). The closure phase for a centrally symmetric brightness distribution is zero. The amount of deviation from zero and the time scales of fluctuations of the closure phases can be used to probe the structure of an asymmetric source even if it cannot be imaged (due to lack of flux or insufficient $uv$ plane coverage). This technique was already used by \cite{Krichbaum2006} and \cite{Lu2011} who found the closure phases of Sgr~A* to be consistent with zero throughout their observations in October 2005 and May 2007, respectively.

We extracted the closure phases from the visibility data for three independent triangles of VLBI stations, initially binned into 10-second time bins. We flagged data with large errors (larger than the standard deviation of the data for a given triangle) and obvious outliers. The fraction of flagged data is 8.9\%, 7.9\%, and 6.0\% for the triangles of EF-PB-PV, EF-PV-YS, and PB-PV-YS, respectively; the difference between the results obtained with and without flagging is insignificant however. We took the weighted average of the remaining values for each scan of 6 minutes; the resulting data set is shown in Fig.~\ref{Closure}. We obtained a combined reduced $\chi^2$ value of 1.34 for all the closure phases for all the independent triangles based on the null hypothesis ``the closure phase is zero all the time''. The corresponding false alarm probability (p-value) is 0.136; this value is clearly too high to reject the null hypothesis (this outcome did not change when choosing bin sizes other than 6 min). Consequentially, we conclude that the closure phases are in agreement with zero during the observing time.

\begin{figure}[!t]
\centering
\includegraphics[trim=12mm 4mm 3mm 9mm, clip, width=85mm]{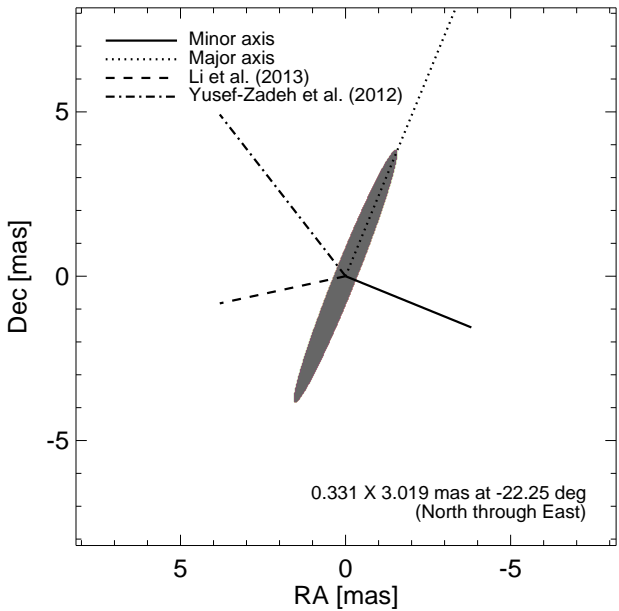}
\caption{Outflow orientations assumed in our simulation. The gray ellipse (with FWHM and position angle as noted) indicates the beam, dashed lines indicate four outflow orientations: along the major axis, along the minor axis, along the jet direction claimed by \cite{Li2013}, and along the one claimed by \cite{Yusef2012}.}
\label{Illustration}
\end{figure}

\section{Discussion \label{sec:discuss}}

The absence of a deviation of the closure phases from zero -- within typical errors of a few degrees -- implies the absence of \emph{asymmetric} structure around Sgr~A* at the time of our observation. A priori, bipolar, symmetric jets can account for zero closure phases. However, if the jet axis is oriented partially along the line of sight, the flux observed from the jet approaching the observer ($f_+$) is amplified while that from the receding jet ($f_-$) is dimmed by the Doppler effect. If the jets on both sides of the black hole have the same luminosity intrinsically, the observed ratio of the two fluxes is
\begin{equation}
\frac{f_+}{f_-} = \left(\frac{1+\beta \cos\theta}{1 - \beta \cos\theta}\right)^{2 + \alpha}
\end{equation}
(e.g., \citealt{Beckmann}), where $\beta$ is the jet speed in units of speed of light, $\theta$ is the angle between the jet axis and the line of sight, and $\alpha$ is the spectral index defined via the flux density $\Sn\propto\nu^{\alpha}$. Various observational constraints ($\alpha\approx0.5$, $\beta\lesssim0.1$; \citealt{Lu2011}) and model predictions for transient jets ($\theta\approx60\degree$, $\beta\approx0.7$; \citealt{Falcke1998, Falcke2009}) place the flux ratios to be expected anywhere between $\approx$1 and $\approx$6. Accordingly, even intrinsically symmetric jets from Sgr~A* might appear asymmetric in the observer frame. 

In order to constrain the size of outflows that we could have missed, we performed simple simulations. We placed artificial, uni-polar secondary sources next to a primary point source model representing Sgr~A* and compared the closure phases obtained from the resulting artificial visibility data with the observations. We considered two geometries: a single point source and a jet composed of 10 equally spaced point sources (knots) with equal fluxes. We probed four orientations for the simulated outflows (See Fig.~\ref{Illustration}): along the major axis, along the minor axis of the beam, the jet direction claimed by \cite{Li2013}, and the jet direction claimed by \cite{Yusef2012}. We used total fluxes of 0.2~Jy and 0.55 Jy for the artificial sources; these values ensure that our simulated outflows are sufficiently faint to not violate the constraints given by the known recent brightness evolution of Sgr~A* (0.2~Jy from the mean variability of $\approx$15\% from June 2013 to February 2014 at 41 GHz, with 0.55~Jy corresponding to the maximum variation in the same period \citep{Chandler2014}). For each simulation setup, we measured the average of absolute values of the closure phases for each triangle. We varied the distances of the model sources (in case of the jet model: the maximum distance) from Sgr~A* until we found a \emph{critical distance} at which the absolute values of the simulated closure phases exceeded those of the observations by more than the $1\sigma$ error at all triangles. We summarized our results in Table~\ref{Model}. Unsurprisingly, critical distances are smaller for brighter outflows. Jet-like structures lead to larger critical distances than equally luminous single, compact sources. As a consequence of the very elongated beam, the critical distances for sources located along the major axis of the beam are larger by a factor of $\approx$7 than for those located along the minor axis. In a few cases (denoted `N/A' in Table~\ref{Model}), the absolute values of the simulated closure phases were comparable to those of the observations for all distances of the model sources, meaning we could not identify a critical distance. Overall, our observations limit the extension of asymmetric (in the observer frame) jet-like outflows from Sgr~A* to projected distances of $\approx$2.5~mas along the major axis and $\approx$0.4~mas along the minor axis.

\begin{table*}[!t]
\centering
\caption{Size constraints for outflows from Sgr~A*.}
\begin{tabular}{llllllllll}
\toprule
    &     & \multicolumn{2}{c}{Minor axis} & \multicolumn{2}{c}{Major axis} & \multicolumn{2}{c}{Li+ 2013} & \multicolumn{2}{c}{YZ+ 2012}\\
\midrule
 & Triangle & ~~0.2 Jy & ~~0.55 Jy & ~~0.2 Jy & ~~0.55 Jy & ~~0.2 Jy & ~~0.55 Jy & ~~0.2 Jy & ~~0.55 Jy\\
\midrule
{\sc Point source} & EF-PB-PV & 0.22 [21] & 0.19 [18] & 1.21 [114] & 1.10 [104] & 0.44 [41] & 0.39 [36] & 0.22 [20] & 0.19 [18] \\
 & EF-PB-YS & 0.22 [21] & 0.19 [18] & 1.85 [174]  & 1.62 [152] & N/A & 0.56 [52] & 0.22 [20] & 0.19 [18] \\
 & PB-PV-YS & 0.24 [22] & 0.21 [20] & 1.66 [156] & 1.51 [142] & 0.59 [55] & 0.47 [44] & 0.23 [22] & 0.21 [20] \\
 & Average  & 0.23 [21] & 0.20 [19] & 1.57 [148] & 1.41 [133] & 0.51 [48] & 0.47 [44] & 0.22 [21] & 0.20 [18] \\
\midrule
{\sc Knotty jet} & EF-PB-PV & 0.36 [34] & 0.28 [26] & 1.84 [173] & 1.52 [143] & 0.67 [63] & 0.54 [51] & 0.35 [33] & 0.27 [25] \\
 & EF-PB-YS & 0.39 [36] & 0.28 [26] & 2.97 [279] & 2.32 [218] & N/A & N/A & 0.37 [34] & 0.27 [26] \\
 & PB-PV-YS & 0.37 [34] & 0.30 [28] & 2.58 [243] & 2.10 [198] & N/A & 0.72 [68] & 0.36 [33] & 0.29 [27] \\
 & Average  & 0.37 [35] & 0.29 [27] & 2.47 [232] & 1.98 [186] & 0.67 [63] & 0.63 [60] & 0.36 [34] & 0.28 [26] \\
\bottomrule
\end{tabular}
\tablefoot{Listed are the critical distances (see Sect.~\ref{sec:discuss} for details) for various cases: four directions for the location of the artificial sources -- along the major axis of the beam, along the minor axis, along the jet direction claimed by \cite{Li2013}, and along the one claimed by \cite{Yusef2012}; two total flux densities of the artificial sources -- 0.2~Jy and 0.55~Jy; and two simple geometries -- a single point source and a knotty jet composed of 10 equidistant point-like components. Values outside [inside] brackets are in units of milliarcseconds [Schwarzschild radii]. All values are projected distances.}
\label{Model}
\end{table*}

Our analysis limits the (projected) extension of linear outflows to about 232 and 35 Schwarzschild radii, respectively, for outflows with fluxes of about 0.2~Jy; obviously, outflows substantially fainter could be larger and still remain undetected. Unfortunately, the resolution of our observations is not sufficient to probe structures in accretion flows expected to occur on scales $\lesssim10\,\RS$ (cf., e.g., \citealt{Broderick2011}); those observations will probably have to await the Event Horizon Telescope (cf. \citealt{Fish2014}). When referring to radio galaxies for comparison, especially to M~87 which has a central black hole with small angular diameter ($\approx$10$\,\mu$as) second only to Sgr~A*, one notes that a distance of tens of Schwarzschild radii appears to be critical for the formation of AGN jets. Recent VLBI observations of the jet of M~87 find a transition in the collimation geometry at a distance of about $100\,\RS$, with the jet opening angle being smaller outside this boundary; this has been interpreted as $\approx$100\,$\RS$ being the characteristic distance for the conversion of magnetic to kinetic energy in a magnetically launched jet \citep{Hada2013}. Accordingly, Sgr~A* is potentially a very important test case for AGN jet physics \emph{if} a jet is ever detected.

Our non-detection of outflows is consistent with earlier null results from VLBI observations of Sgr~A* at 86~GHz \citep{Krichbaum1998, Lo1998, Shen2005, Lu2011} despite the fact that we observed Sgr~A* at an epoch of potentially increased accretion. Our results are in line with other recent observations finding that Sgr~A* has been quiescent from radio to X-rays in 2013 and 2014 \citep{Akiyama2013, Brunthaler2013, Chandler2014, Degenaar2014}. In addition, the zero closure phases within errors can put some constraints on the substructure in scattering disks of Sgr A* (e.g., \citealt{Gwinn2014}); this substructure must remain symmetric on spatial scales of from sub-mas to a few mas depending on an axis in the geometry of the substructure. 

As noted above, the radio flux density that can realistically be attributed to accretion of parts of G2 is $\Sn\approx0.2$~Jy, translating into a radio luminosity $\LR=4\pi R_0^2\nu\Sn\approx1.3\times10^{26}~{\rm W}\approx0.3\,L_{\odot}$ (for $\nu=86$~GHz). Without making further assumptions, we can state that this value corresponds to an effective accretion rate $\eta\dot{M}=\LR/c^2\approx1.5\times10^9~{\rm kg\,s}^{-1}\approx7.7\times10^{-9}\,M_{\oplus}\,{\rm yr}^{-1}$, with $\eta\in[0,1]$ being the matter-to-light conversion efficiency and $c$ denoting the speed of light. However, in addition we have to take into account that accreted matter might not be converted into electromagnetic radiation but into jets, with the presence of jets being an \emph{ad hoc} working hypothesis. In this case, a rough quantitative estimate of the accretion rate is provided by the jet power--accretion power relation of radio galaxies: for highly sub-Eddington accretion (as is the case for Sgr~A*), kinetic jet power $\Pj$ and accretion power $\Pacc\equiv\dot{M}c^2$ are related like $\Pj\approx0.01\Pacc$ (e.g., \citealt{Allen2006,Trippe2014}). The powers $\Pj$ and $\LR$ in turn are empirically related like $\Pj\approx5.8\times10^{36}\,(\LR/10^{33})^{0.7}$~W \citep{Cavagnolo2010}. Combining the two relations and using, again, $\LR\lesssim0.2$~Jy implies an accretion rate $\dot{M}\lesssim10^{17}\,{\rm kg\,s}^{-1}\approx0.5\,M_{\oplus}\,{\rm yr}^{-1}$. We note that this calculation assumes a highly idealized situation, neglecting interactions between G2 and the accretion flow around Sgr~A*. Given the small accretion rates as well as the complexity of the accretion flows, it seems realistic that relatively large amounts of matter could be ``peeled off'' G2 and driven out of the accretion zone by winds or other non-collimated outflows.

These limits on accretion rates, which are always substantially below the total mass of G2 (with details depending on which structure of G2 is assumed), are consistent with the observed kinematics of G2 during its pericenter passage: as noted by several studies, the orbit was purely Keplerian even after the pericenter passage \citep{Witzel2014, Pfuhl2015, Valencia-S.2015}. This indicates that G2 did not experience a notable loss of angular momentum and energy, pointing at rather weak interactions with the hot gas around Sgr~A*. As suggested by several studies based on hydrodynamical simulations, the viscous timescale can be on the order of years \citep{Burkert2012, Schartmann2012, Moscibrodzka2012}, meaning it could take a few more years to see AGN-like (mostly in radio mode associated with hot accretion flows and jets, see \citealt{Yuan2014} for a review) activity in Sgr~A*.

\section{Summary and conclusions}

We observed Sgr~A* with four European GMVA stations at 86 GHz on 1 October 2013, searching for centrally asymmetric (in the observer frame) outflows potentially generated as a consequence of enhanced accretion during the pericenter passage of the gas cloud G2. Our key findings are:

\begin{enumerate}

\item  The closure phases measured for Sgr~A* remained zero during the observation time, corresponding to a non-detection (within errors) of asymmetric structure. We are able to constrain the size of the outflows that we could have missed to $\approx$2.5 mas along the major axis and $\approx$0.4 mas along the minor axis of the beam, corresponding to approximately 232 and 35 Schwarzschild radii, respectively. These sizes are on the order of the scale ($\approx$100\,$\RS$) where magnetically launched AGN jets are supposed to convert magnetic into kinetic energy, but are not yet sufficient to probe structures on scales $\lesssim$10\,$\RS$ expected in hot accretion flows. 

\item  As probably less than 0.2~Jy of the flux from Sgr~A* can be attributed to extra accretion from G2, one finds an effective accretion rate $\eta\dot{M}\lesssim1.5\times10^9~{\rm kg\,s}^{-1}\approx7.7\times10^{-9}\,M_{\oplus}\,{\rm yr}^{-1}$ for material from G2. Exploiting the kinetic jet power--accretion power relation of radio galaxies, one finds that the accretion of matter that ends up in jets is limited to $\dot{M}\lesssim10^{17}\,{\rm kg\,s}^{-1}\approx0.5M_{\oplus}\,{\rm yr}^{-1}$.

\end{enumerate}

Overall, our analysis suggests that G2 is largely stable against loss of angular momentum and subsequent (partial) accretion at least on time scales $\lesssim$1 year.

\begin{acknowledgements}
We are grateful to {\sc Eric W. Greisen} (NRAO, U.S.A.) for technical support. We acknowledge financial support from the Korean National Research Foundation (NRF) via Global PhD Fellowship Grant 2014H1A2A1018695 (J.-H. P.) and Basic Research Grant 2012R1A1A2041387 (S. T.). We made use of data obtained with the Global Millimeter VLBI Array (GMVA), which consists of telescopes operated by the MPIfR, IRAM, Onsala, Metsahovi, Yebes and the VLBA. The data were correlated at the correlator of the MPIfR in Bonn, Germany. Last but not least, we are gratful to the anonymous referee for a helpful review and valuable suggestions.
\end{acknowledgements}

\end{document}